\documentclass[twocolumn]{revtex4}

\usepackage{amsmath, amsthm, amssymb}
\usepackage{graphicx}
\newtheorem{theorem}{Theorem}[section]
\newtheorem{proposition}{Proposition}  
\newtheorem{corollary}{Corollary}	
\numberwithin{equation}{section}
\numberwithin{theorem}{section}
\numberwithin{proposition}{section}
\numberwithin{corollary}{section}

\begin{document}

\title{The Koide Lepton Mass Formula and Geometry of Circles}


\author{Jerzy Kocik}
\affiliation{Department of Mathematics, Southern Illinois University, Carbondale, IL 62901}

\email{jkocik@siu.edu}


\begin{abstract}
A remarkable formal similarity between Koide's Lepton mass formula and a generalized Descartes circle formula is reported.
\end{abstract}



\maketitle

\section{Introduction}

Quarks and leptons are believed to be the fundamental particles of matter yet their nature is still far from being understood. 
One of the exciting puzzles is a formula involving the masses of the three leptons, discovered 
by Yoshio Koide \cite{YK1, YK2}:
\begin{equation}    \label{eq:1}
  m_e + m_\mu + m_\tau = \dfrac{2}{3} 
	(\surd m_e + \surd m_\mu + \surd m_\tau)^2.
\end{equation}
See Table 1 for the corresponding numerical values (from \cite{Nak}).

\begin{table}[h!]		\label{tbl:1}
\begin{tabular}{|lll|ll|}	\hline 
  &  \qquad &&& \\[-10pt]
  &electron& &&$\phantom{\surd}m_e = 0.510998910$ MeV/c${}^2$ 	\\
  &	   & &&\hphantom{$\surd m_e$} $= 1\,m_e$	\\ 
  &	   & &&$\surd m_e = 1$  \\[4pt]
  &muon&     &&$\phantom{\surd}m_\mu = 105.658367$ MeV/c${}^2$  	\\
  &	   & &&\hphantom{$\surd m_e$} $= 206.768282 \ m_e$ \\
  &  	   & &&$\surd m_{\mu} = 14.37943957 \ m^{1/2}_e$ \\  [4pt]
  &tau    & &&$\phantom{\surd}m_\tau = 1776.84$ MeV/c${}^2$   \\
  &	   & &&\hphantom{$\surd m_e$} $= 3477.1894\ m_e$	\\
  &	   & && $\surd m_\tau = 58.97 $ 
		\hbox to .1in{} \\[6pt] \hline
\end{tabular}
\caption{Lepton masses.}
\end{table}

Quite remarkably, Koide used his formula to predict the mass of the tau lepton with surprising accuracy:  

\begin{center}
\begin{tabular}{lll}
  old measured	&$m_\tau = 1784 \pm 4$ 	 &[1970's] \\
  predicted by Koide \  &$m_\tau = 1776.97$     &[1982]	\\
  new measured	&$m_\tau = 1776.99 \pm .3$    &[2002]	\\
  newer measured &$m_\tau = 1776.84 \pm .17$ \quad &[2011]
\end{tabular}
\end{center}

\noindent
(all in MeV/c${}^2$).  Some interesting formal associations have been noticed since
\cite{Bra,GG,LM,RG},      
but after almost 30 years the consensus is that the ``mystery of the lepton mass formula''  \cite{YK3} 
remains unsolved.

\section{From Descartes to Koide}

\begin{figure}[h!] %
  \centering
  \includegraphics[width=3.3in,keepaspectratio]{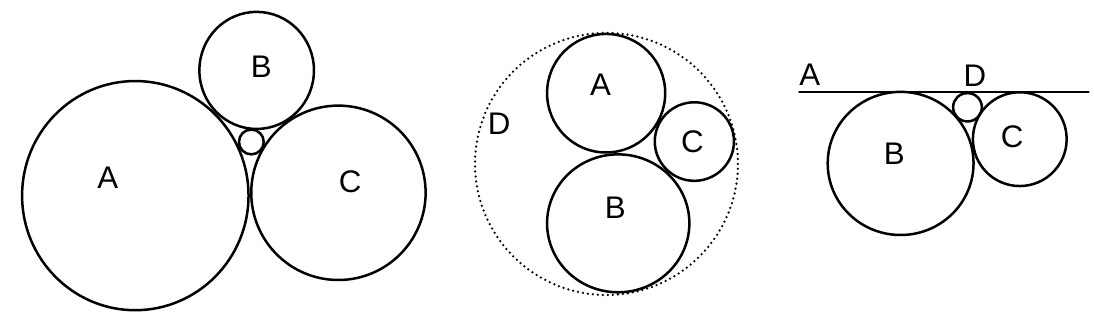}
  \caption{Four circle in various Descartes configurations.}
  \label{fig:fig-1}
\end{figure}

In this note we want to call attention to a curious formal similarity of Koide's formula to Descartes's circle formula.  Descartes -- in his 1654 letter to the princess of Bohemia, Elizabeth II -- showed that the curvatures of four mutually tangent circles (reciprocal of radii), say \textit{a,b,c,d}, satisfy the following ``Descartes's formula'':

\begin{theorem}[Descartes's Circle Formula, 1654 \cite{Des}]  The curvatures of four circles in Descartes's configuration satisfy this equation:
\begin{equation}    \label{eq:2}
  (a+b+c+d)^2 = 2(a^2+b^2+c^2+d^2)
\end{equation}
where $a =1/r_1$, $b = 1/r_2$, etc., denote {\normalfont reciprocals} of radii called {\normalfont bends}, which are signed {\normalfont curvatures}. In {\normalfont Fig.~\ref{fig:fig-1}} in the middle, circle $D$ is a boundary of an unbounded disk (region outside $D$) hence its bend is negative.  Circle $D$ in the right figure has bend equal zero.
\end{theorem}
%

 The similarity of Descartes's circle formula to Koide's lepton mass formula is striking:  represent masses as reciprocals of disk areas, equivalently, their curvatures as square roots of masses, and we get \eqref{eq:2} except for a different coefficient for the sum of squares: 2 instead of 2/3.   The fourth term may be taken as zero, $d = 0$, corresponding geometrically to a line as a special circle.  

The factor of 2/3 invalidates the exact correspondence with Descartes's formula.  
However, the present author has found a formula for circles in general position that generalizes Descartes's.  

\begin{theorem}[General Circle Formula \cite{JK}]  
The radii of $n+2(n-1)$-spheres $C_1,\ldots,C_{n+2}$  in general position in $R^n$ satisfy a quadratic equation
\begin{equation}    \label{eq:3}
  BFB^T = 0
\end{equation}
or simply
\[
  F_{ij} b_ib_j = 0
\]
where $B = [b_1,\ldots,b_n]$ is the row matrix (covector) of the curvatures of the spheres and $F = f^{-1}$ is the inverse of the ``configuration matrix'' $f$ with entries defined by
\begin{equation}    \label{eq:4}
  f_{ij} = \begin{cases}
	(d_{ij} - r_i - r_j)/2r_ir_j	&\text{in general}  \\
	\cos \varphi,		&\text{if $C_i$ and $C_j$ intersect.}
	\end{cases}
\end{equation}
\end{theorem}
Here $d_{ij} =$ distance between the centers of $i^{th}$ and $j^{th}$ spheres, $r_{i} =$ radius of the $i^{th}$ sphere. For spheres that intersect,  $\varphi$ is the angle of their intersection.  
(Quite interestingly, the above theorem can be derived from the fact that circles in the Euclidean plane 
may be regarded as vectors in the Minkowski space \cite{JK}.)

Consider the special case where the product for every pair is the same, say $p$.  Then one calculates the inverse of the corresponding matrix $f$:
\[
  \begin{aligned}
  f &= \begin{bmatrix}
	-1 &p &p &p	\\
	p &-1 &p &p 	\\
	p &p &-1 &p 	\\
	p &p &p &-1	
	\end{bmatrix} 	\\
  \Rightarrow F &= \dfrac{1}{3p^2+2p-1}
	\begin{bmatrix}
	1-2 &p &p &p	\\
	p &1-2 &p &p 	\\
	p &p &1-2 &p 	\\
	p &p &p &1-2	
	\end{bmatrix} 
  \end{aligned}
\]
Solving \eqref{eq:3} gives the following geometric result:

\begin{proposition} 
  Four circles of curvatures $a$, $b$, $c$, $d$, respectively, intersecting pairwise at the angle $\varphi = \arccos p$ satisfy the quadratic equation
\begin{equation}    \label{eq:5}
  (a^2+b^2+c^2+d^2) = \dfrac{p}{3p-1} (a+b+c+d)^2.
\end{equation}
\end{proposition}

For circles mutually tangent we have \textit{p} = 1 and the above formula reduces to Descartes's formula.  We are, however, interested in the value 2/3 for the coefficient on the right side.  

\begin{corollary}
Koide's lepton mass formula may be interpreted as a Descartes-like circle formula \eqref{eq:3} for $p = 2/3$, where the squared curvatures correspond to the masses of the three leptons.  One mass is assumed zero and corresponds to a straight line.  Each pair of circles intersects under the same angle:  
\[
  \varphi = \arccos(2/3) \approx 0.841 \hbox{ rad } \approx
	15/56\, \pi \approx 48.2^{\circ}.
\]
Interestingly, the radii, reciprocals of square roots of masses, are close to integer values, namely,  $r_\tau = 1$,  $r_\mu \approx 4.10$,  $r_e \approx 58.97$,  but not close enough to warrant further attention.
\end{corollary}

\begin{figure}[h!] %
  \centering
  \includegraphics[width=2in]{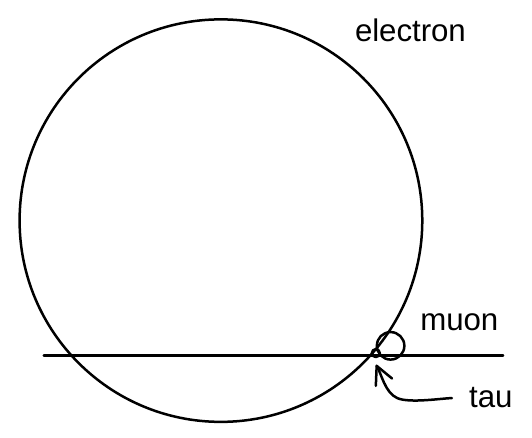} 

  \includegraphics[width=2in]{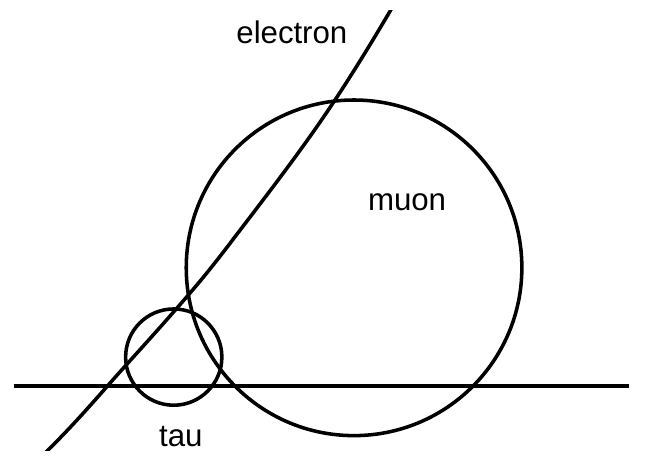}
  \label{fig:fig-2}
  \caption{Lepton masses via circle configuration (detail on bottom).}
\end{figure}

Since the Koide formula might be applicable for other particle families, let us provide a solution to the configuration of $n+2$ spheres in $n$-dimensional space: 
\begin{equation}    \label{eq:6}
  b^2_1 + b^2_2 + \ldots + b^2_{n+2}
	= \dfrac{p}{(n+1)p-1} (b_1 + \ldots + b_{n+2})^2
\end{equation}
that may be a candidate for geometry of generalized Koide formula:
\[
  m_1 + m_2 + m_3 + \ldots
	= \kappa \cdot \left( \sqrt {m_1} + \sqrt {m_2}
		+ \sqrt {m_3} + \ldots \right)^2.
\]
The relation is thus 

\begin{equation}    \label{eq:7}
  p = \dfrac{1}{n+1-1/\kappa} \qquad \hbox{and} \qquad
	\kappa = \dfrac{2}{3} \Rightarrow p = \dfrac{2}{2n-1}\,.
\end{equation}
%

\section{Conclusions}

We conclude with some remarks:

\begin{description}
\item{\textbf{1. Leptons and segments.}}
If one considers three segments on a line rather than circles in plane, formula \eqref{eq:3} with $n=1$, becomes 
\[
  a^2+b^2+c^2 = \dfrac{p}{2p-1}(a+b+c)^2.
\]
To make the coefficient in Koide's formula equal $\kappa = 2/3$ we would need $p = 2$. The segments for $e$, $\mu$, $\tau$ would be $(-1.40, 1.40)$,  $(-1.49, 1.68)$, $(-1.42, 1.47)$, respectively. 

\item{\textbf{2.  Neutrinos.}} 
Recently, Brannen observed \cite{Bra} that the same pattern is followed by the neutrino masses under the condition that the square root of the first mass is taken to be negative.  


With his data,  $m_1= 0.000388eV$, $m_2= 0.00895eV$, $m_3= 0.0507eV$,
the factor of 2/3 is obtained with the very good precision of 0.01\% and the resulting equation 
\[
  m_1 + m_2 + m_3 = \dfrac{2}{3} \left( -\surd m_1 + \surd m_2 
	+ \surd m_3 \right)^2
\]
poses no problem to our geometric interpretation, as negative curvature is associated to unbounded disk defined by the region outside a circle.
\begin{figure}[h!] %
  \centering
  \includegraphics[width=1.2in]{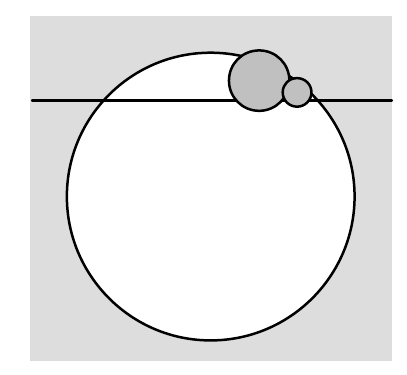} 
  \includegraphics[width=1.75in]{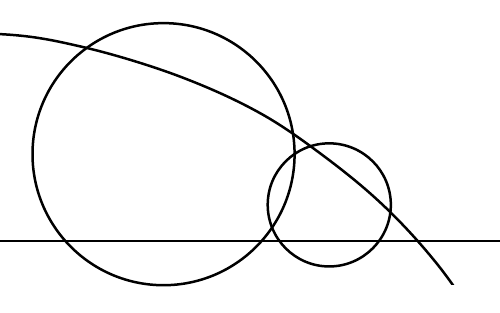}
  \label{fig:fig-3}
  \caption{Disk configuration for a negative $\sqrt{m}$.}
\end{figure}

But estimations of neutrino masses change rapidly with new experiments and only upper bounds are measured.
Most recent estimations are:  electron neutrino: $m < 2.2 eV$,  
muon neutrino: $m < 170$ keV, and tau neutrino: $m < 15.5$ MeV   \cite{CUPP}.
Thus,  taking the highest value for the muon neutrino would suggest a rather low value of 2.4 MeV for 
the tau neutrino if Koide's formula with ratio $2/3$ were to be satisfied.

\item{\textbf{3.  Quarks.}} 
Quark mass estimations taken from \cite{RG} are presented in the table below.
An extended version of Koide's equation would suggest 2/3 with a precision of 5\% \cite{RG}.  The geometry here would be that of six 4-spheres in a five-dimensional space with $p = 2/9$, giving an intersection angle of $\varphi \approx (3/7)\pi \approx 77.2^\circ$.  We note that the precision grows rapidly to 0.1\% when only the last three quarks, $c$, $b$ and $t$, are considered.

\begin{table}[h!]
\begin{tabular}{|lll|lll|ll|} \hline
  &\quad &&&&&& \\[-10pt]
  &up	&&&$m_u = 1 m_u$	&&&$\surd m_u = 1$	\\
  &down	&&&$m_d = 12 m_u$	&&&$\surd m_d = 3.464$	\\
  &strange &&&$m_s = 210 m_u$	&&&$\surd m_s = 14.491$	\\
  &charm &&&$m_c = 2500 m_u$	&&&$\surd m_c = 50$	\\
  &bottom&&&$m_b = 9000 m_u$	&&&$\surd m_b = 94.868$	\\
  &top	&&&$m_t = 348000 m_u$	&&&$\surd m_t = 589.915$ \qquad	\\[6pt]\hline
\end{tabular}
\caption{Quark masses}		\label{tbl:3}
\end{table}

\item{\textbf{4. Coda.}}
Whether this intriguing connection with geometry will contribute to an understanding of the masses of leptons remains an interesting question and requires further investigation.  The analogy described above may turn out to be merely superficial, but given the current state of understanding about the matter, any interesting structural parallels are worthy of our consideration in the effort of reconstructing deeper patterns.

\end{description}

\vskip2em

\section*{Acknowledgements}
I am grateful to Philip Feinsilver for his encouraging interest and priceless comments.

\end{document}